# AI-Enhanced Diagnosis of Peyronie's Disease: A Novel Approach Using Computer Vision

**Yudara Kularathne, Janitha Prathapa, Prarththanan Sothyrajah, Salomi Arasaratnam, Sithira Ambepitiya, Thanveer Ahamed, Dinuka Wijesundara**

HeHealth Inc, USA

{yu, prathapa, prarththanan, salomi, sithira, thanveer, dinuka}@hehealth.ai

## Abstract

This study presents an innovative new AI-driven tool for diagnosing Peyronie's Disease (PD), which affects 0.3% – 13.1% men worldwide. Our method leverages key point detection on both images and videos to measure penile curvature angles among advanced computer vision techniques. This tool has shown high accuracy identifying anatomical landmark when validated against conventional goniometer measurements. Traditional PD diagnosis often involves subjective and invasive methods which may result in patient discomfort and inaccuracies. Our approach offers a precise, reliable and non-invasive diagnostic tool to address these drawbacks. The model differentiates between PD and normal anatomical changes with the sensitivity of 96.7% and specificity of 100%. This breakthrough marks a significant improvement in urological diagnostics by greatly enhancing the efficacy and convenience of PD assessment for both healthcare providers and patients.

## 1 Introduction

The Peyronie's Disease (PD) is defined by the growth of fibrous plaques in the penis which lead to curvature and discomfort during erections. The exact cause of PD is yet unknown. According to global data, 0.3% - 13.1% of men suffer from PD. Among them, greater prevalence of 16% is seen in specific groups who undergo radical prostatectomy [1]. Further, screening research on prostate cancer in United States revealed that 8.9% prevalence of PD was found among the participants [2]. The prevalence of PD has been further shown by a number of other surveys and research studies carried out in the past year [3, 4]. These results from the studies suggest that the prevalence of PD may actually higher than prior estimates, stressing the need for the precise diagnostic system in addition to proper accurate reporting and increased global awareness.

To diagnose PD, urologist - the primary healthcare professionals diagnosing Peyronie's disease – usually rely on physical examination to look for a palpable penile plaque [2]. Other techniques such as Intracavernous injection and stimulation [5], and ultrasonography [6] may be included. In addition, patients may be requested to provide photos of their erect penis taken from different angles for diagnostic purposes. But these traditional methods have number of limitations. First, urologists may have different interpretation for same condition as the physical examinations are subjective. Furthermore, a natural erection may not be same as the penile erection observed in the clinical settings, which could lead to incorrect diagnosis. In addition, traditional methods often fail in early detection, and can make patient feel discomfort and embarrassment.

Recent advancement in automated deep learning technologies have shown great potential for enhancing diagnosis accuracy and customizing treatment strategies. This advancement could significantly improve patient outcome, particularly when applied to the diagnosis of PD. Our project uses a computer vision-based system that analyze 2D photos and videos for comprehensive diagnosis, representing a groundbreaking innovation determining the degree of curvature. Compared to conventional image-based evaluations, our video-based approach offers a significant improvement and a more comprehensive and nuanced knowledge of the situation.

In conclusion, development of our innovative penile curvature assessment tool represents a significant breakthrough in the detection and surgical treatment of Peyronie's disease (PD). It greatly improves surgeons' ability to make accurate decisions by providing them with real-time quantification of the degree of penile curvature throughout the procedure during surgery.

## 2 Background and Related works

Peyronies Disease is a condition that results in painful, curved erections, as a result of the growth of fibrous scar tissue inside the penis. Although the specific cause of PD is not yet known, it is believed to be as a result of an autoimmune response or repeated penile injury. The Key symptom of the PD is the penile curvature. Accurately determining the degree of angle is essential for treatment planning. There are several methods used for diagnosis of PD: The most accurate method for measuring penile curvature in Peyronie's disease (PD) is the in-office goniometric angle measurement of a pharmacological induced erection [7]. According to [7], goniometer is considered the golden standard for evaluating penile curvature due to its accuracy and reliability. Also penile at-home auto photography is used for measuring penile curvature, which underestimate the degree of penile curvature due to the lack of standardized procedures and potential variations in self-photography technique [8].

Given the limitation and variability associated with traditional penile curvature assessment methods in PD, recent advancements in AI have potential to address and standardize the accuracy of penile curvature assessment. There are some previous studies that have been demonstrated the application of AI tools for penile curvature assessment. Notably, a pilot study in [9] used AI for penile curvature assessment, employing a Hough-Transform-based angle estimation algorithm. However, this approach has notable limitations, particularly its lack of robustness in many cases. Subsequent study [10] has addressed these short coming, by applying a deep learning model for angle calculation. There are some limitations for this approach, as it will be affected by camera angle and lighting conditions. Additionally, all the previous study has been conducted on 3D penile models, rather than real images. To overcome these limitations, here we developed a novel approach to calculate the angle using a keypoint detection model using a robust state-of-the-art deep learning model. In our approach, we utilize video input to achieve a more robust and precise calculation of angles, rather than solely relying on static images. To the best of our knowledge, this is the first study conducted using real penile images and represents the first real-world applicable model for the diagnosis of Peyronie's Disease (PD).

## 3 Methodology

### 3.1 Data Collection

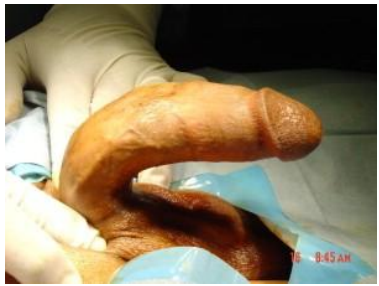
(a) Real PD penis image captured in surgery.

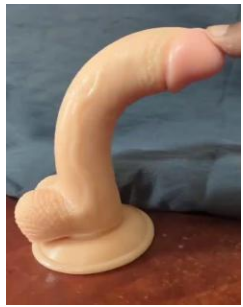
(b) Skin-colored, penis-shaped sex toy.

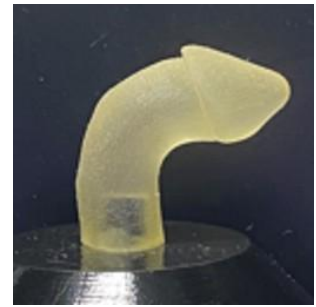
(c) 3D printed penile model.

Figure 1: Sample of the dataset we have collected.

For the model training, we have collected diverse images and video frames consisting of a wide range of penile curvatures. For the training we have used images from different resources, which are listed below.

- We have utilized a curated set of 200 real images that we have collected, showcasing a range of penile curvature degrees.
- A set of 50 images of 3D printed penile models [9]. Each model measures approximately 1.5 cm in width and 5-6 cm in length, featuring different uniplanar hinging curvature angles. This inclusion aims to provide a controlled set of curvature examples for the model.

- Around 100 frames extracted from videos featuring skin-colored, penis-shaped sex toys with varying curvatures. These frames provide dynamic representation of curvature, improving the model's ability to interpret and analyze angle changes and movements.

### 3.2 Data Annotation

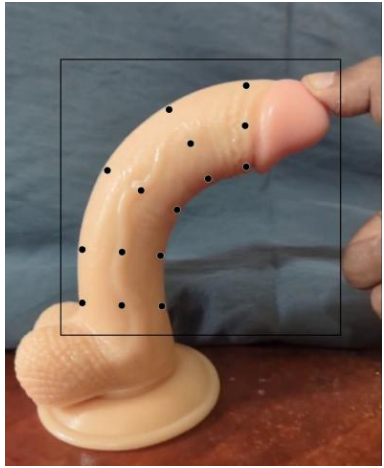

Figure 2: Example annotation.

A free and open-source image and video annotation tool called CVAT.ai was used to annotate our picture dataset. Our approach included key point and bounding box annotation into a single XML file. A python script was then used to convert this file into individual text files for each picture, preparing the data in the specific format needed for training the YOLO (You Only Look Once) model.

**["class_id x_center y_center width height x1 y1 x2 y2 ... xn yn"]**

- **'class_id':** An integer representing the class of the object.
- **'x_center y_center':** The x and y coordinates of the center of the bounding box, normalized by the width and height of the image (values between 0 and 1).
- **'width and height':** The width and height of the bounding box, also normalized to values between 0 and 1.
- **'x1 y1, x2 y2, ..., xn yn':** are the normalized x and y coordinates of each keypoint. The number of key points (n) depends on the specific requirements of the task. In our case n =15

### 3.3 System Architecture

We have developed an innovative method in our studies to estimate penile curvature from two-dimensional pictures. This approach utilizes a detection system comprising 15 key points, arranged in three rows with five points each, positioned along the central and lateral aspects of the penis image as shown in the Figure 2. Additionally, our methodology integrates abounding box detection mechanism, specifically designed to accurately localize the penis within the image. For the underlying architecture, we have chosen YoloV8-nano (YoloV8n) [11], a cutting-edge model in the YOLO series. Our tailored YoloV8n model was meticulously trained using a data set of 350 images, as mentioned in the 'Data Collection' section of our study. Our system architecture clearly shown in the Figure 4. Initially, input images are fed into the trained YoloV8 model, which generates key points. These key points are then inputted into the angle calculation module, as detailed in the Section 3.3.3.

Our method utilized video inputs instead of relying only on static 2D images to improve the accuracy and liability of our predictive analysis. This strategic decision based on the understanding that because images are two-dimensional, they can be greatly affected by changes in camera angle and placement [12], which may result in inaccurate measurements of penile curvature. Videos, on the other hand, provide a more dynamic and comprehensive perspective. They make it possible to examine the penile structure from various angles and in various motion states, offering a more comprehensive set of data from which to more precisely determine the degree and properties of curvature.

Therefore, compared to conventional image-based assessments, our video-based approach offers a significant improvement and a more comprehensive and nuanced understanding of the situation. This development is essential to our efforts to transform the Peyronie's disease diagnosis procedure by providing urologists with a tool that is more accurate and more representative of the three-dimensional reality of the illness. To find the most suitable architecture for our application, we have tested two different architectures.

### 3.3.1 Architecture 1

Using YOLOv8n, we trained a key point detection model in the first architecture to find 15 key points that were strategically positioned throughout the penis image. These key points were chosen in order to capture important anatomical landmarks that were necessary for our analysis. As explained in Section 3.3.3, an angle estimation module was used to estimate particular angles between the key points after they were identified. Precise tracking and analysis of positional relationships within the image are made possible by this two-stage method, which consists of key point identification and angle estimates. Figure 3 shows the overall architecture of the system .

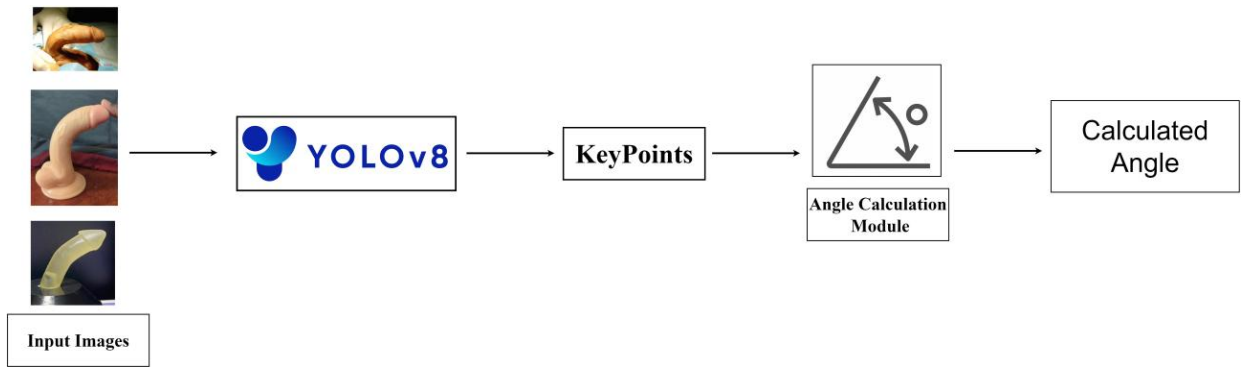

Figure 3: System Architecture-1.

### 3.3.2 Architecture 2

In the second architecture, we separated the penis from the background image using a segmentation model. The penis' boundaries are precisely drawn by the segmentation model, which also isolates it from surrounding background objects in the image. Following segmentation, a keypoint detection algorithm processes the isolated area, identifying 15 key points along the segmented penis that are positioned to capture essential anatomical landmarks. Then. An angle estimate module determined precise angles between the identified key points. This architecture improves precision by working on a clearer, more focused image by combining segmentation with key point identification and angle computation. Figure 4 shows the overall architecture of the system.

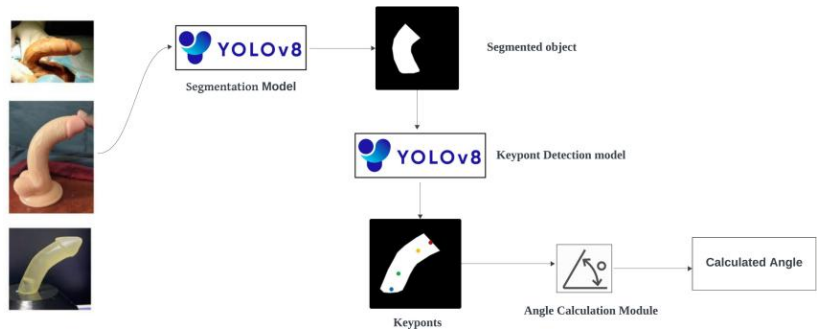

Figure 4: System Architecture-2.

### 3.3.3 Angle Calculation Module

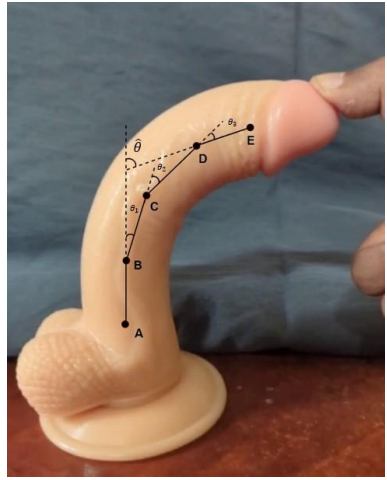

Figure 5: Image showing angle calculation.

Our model is meticulously trained to identify 15 key points, strategically distributed across the penis image. Although the angle of penile curvature could potentially be calculated using just three points on each line, our approach deliberately employs five points per line. This method significantly improves the accuracy in measuring the deviation of the penis tip from its base, particularly in instances where the curvature is not centrally located. Moreover, this five-point per line strategy is crucial in precisely pinpointing the specific location of the curvature.

The angles between predicted key points in our model are determined using an angle calculation module, which employs the Dot Product or Cosine Formula method. Algorithm is explained below,

---
**Algorithm 1** Pseudocode for Angle Calculation module

**Data:** Points A($A_x$, $A_y$), B($B_x$, $B_y$), C($C_x$, $C_y$), and D($D_x$, $D_y$)
**Result:** Angle between vectors ($\overrightarrow{AB}$) and ($\overrightarrow{CD}$) in degrees
**Vector representation:** $\overrightarrow{AB} \leftarrow (B_x - A_x, B_y - A_y)$, $\overrightarrow{CD} \leftarrow (D_x - C_x, D_y - C_y)$
**Dot Product:** $dotProduct \leftarrow (B_x - A_x) \times (D_x - C_x) + (B_y - A_y) \times (D_y - C_y)$
**Magnitude calculation:** $|AB| \leftarrow \sqrt{(B_x - A_x)^2 + (B_y - A_y)^2}$, $|CD| \leftarrow \sqrt{(D_x - C_x)^2 + (D_y - C_y)^2}$
**Angle calculation:** $\vartheta \leftarrow \arccos \frac{dotProduct}{|AB| \times |CD|}$
**Conversion in degrees:** $angle \leftarrow \vartheta \times \frac{180}{\pi}$
**Return:** $angle$

---

We calculate two distinct types of angles for comprehensive analysis. The first angle ($\hat{\vartheta}$) measures the degree of deviation of the penis tip from the base and the second angle is determined at specific points to assess the curvature of the penis ($\vartheta_1$, $\vartheta_2$ and $\vartheta_3$) as shown the figure 5. We calculate four angles along a single line of five key points, which consists of one deviation angle and three angles for curvature assessment. As there are three lines of key points for analysis, we consider only the middle line for the angle. Also angle in the video changes with camera deviation, therefore we consider only the maximum angle as the actual curvature angle.

## 4 Results

In our study, we evaluated the effectiveness of our computer vision approach for diagnosing Peyronie's disease using a dataset of 60 penile images. The dataset was evenly divided, comprising 30 images of patients diagnosed with Peyronie's disease and 30 images of individuals without the condition.

To assess the performance of our technique, we employed several key metrics, namely accuracy, sensitivity, and specificity. These metrics are defined by the following equations:

$$\text{Accuracy (A)} = \frac{TP + TN}{TP + TN + FP + FN} \quad (1)$$

$$\text{Sensitivity} = \frac{TP}{TP + FN} \quad (2)$$

$$\text{Specificity} = \frac{TN}{TN + FP} \quad (3)$$

### 4.1 Results of Architecture 1

Architecture 2 misclassified several images in both the Peyronie's disease and normal categories. It correctly identified only 21 out of the 30 images in the Peyronie's disease category, resulting in a sensitivity rate of 70.0%. In the normal category, it misclassified 5 images as Peyronie's disease, yielding a specificity rate of 83.3%. Despite efforts to enhance the system's accuracy, this architecture struggled to consistently differentiate between Peyronie's disease and normal anatomical variations. For this study, we considered an angle of more than 30 degrees as indicative of Peyronie's disease, but the segmentation process likely introduced errors that impacted overall performance.

Table 1: Classification Results of Architecture 2 for Peyronie's Disease Diagnosis

| N = 60 | Actual: Peyronie's | Actual: Normal |
|---|---|---|
| **Detected: Peyronie's** | 21 (TP) | 5 (FP) |
| **Detected: Normal** | 9 (FN) | 25 (TN) |

The performance of Architecture 2 for Peyronie's disease diagnosis is detailed in Table 1, where misclassifications are evident, significantly lowering its diagnostic reliability.

Table 2: Performance results of Architecture 2

| Criteria | Score |
|---|---|
| Accuracy | 0.77 |
| Sensitivity | 0.70 |
| Specificity | 0.83 |

The performance of the YOLOv8 model for diagnosing Peyronie's disease within this architecture showed suboptimal results, particularly in keypoint detection. The mean average precision at a 50% IoU threshold (mAP@50) was 0.751, highlighting a decrease in the model's accuracy for identifying anatomical landmarks. The mean average precision across IoU thresholds from 50% to 95% (mAP@50-95) was lower at 0.605, indicating further inconsistency in anatomical pose estimation. These findings suggest that incorporating segmentation into this architecture impacted its overall accuracy and reliability.

### 4.2 Results of Architecture 2

Architecture 2 achieved a sensitivity rate of 96.7% by correctly classifying 29 out of 30 images in the Peyronie's disease category. In the normal category, all 30 images were accurately identified, resulting in a specificity rate of 100%. For this study, an angle greater than 30 degrees was used as the threshold for diagnosing Peyronie's disease, in accordance with clinical guidelines. These results demonstrate the high accuracy and reliability of our approach in distinguishing between Peyronie's disease and normal anatomical variations.

Table 3: Classification Results of Our Approach for Peyronie's Disease Diagnosis

| N = 60 | Actual: Peyronie's | Actual: Normal |
|---|---|---|
| **Detected: Peyronie's** | 29 (TP) | 0 (FP) |
| **Detected: Normal** | 1 (FN) | 30 (TN) |

The performance of our diagnostic approach for Peyronie's disease is detailed in Table 1, according to the evaluation criteria utilized.

Table 4: Performance results of our appraoch

| Criteria | Score |
|---|---|
| Accuracy | 0.98 |
| Sensitivity | 0.97 |
| Specificity | 1.00 |

All the Yolov8 model's performance in diagnosing Peyronie's disease was assessed using keypoint detection metrics. It achieved a mean average precision at a 50% IoU threshold (mAP@50) of 0.995, indicating high accuracy in identifying anatomical landmarks. The mean average precision across IoU thresholds from 50% to 95% (mAP@50-95) was 0.944, demonstrating robust performance in anatomical pose estimation. These results highlight the model's precision and reliability in clinical applications for Peyronie's disease. As real images are sensitive, we have provided examples using toy images for illustration in Figure6

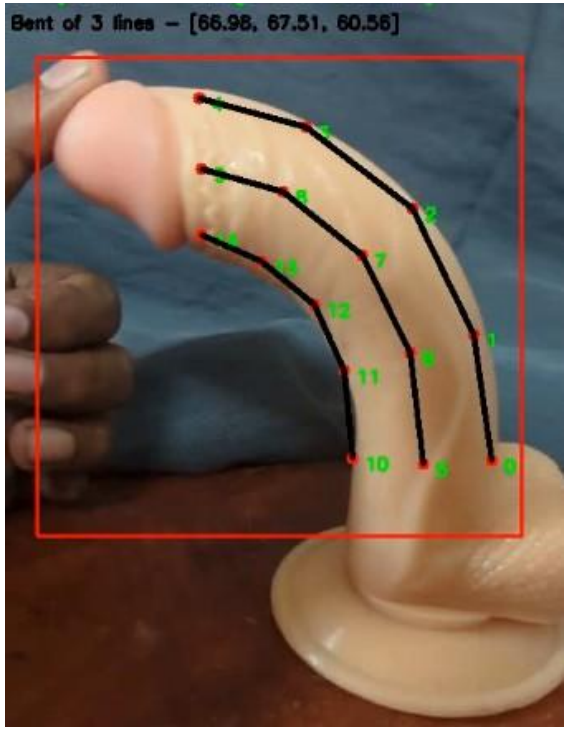
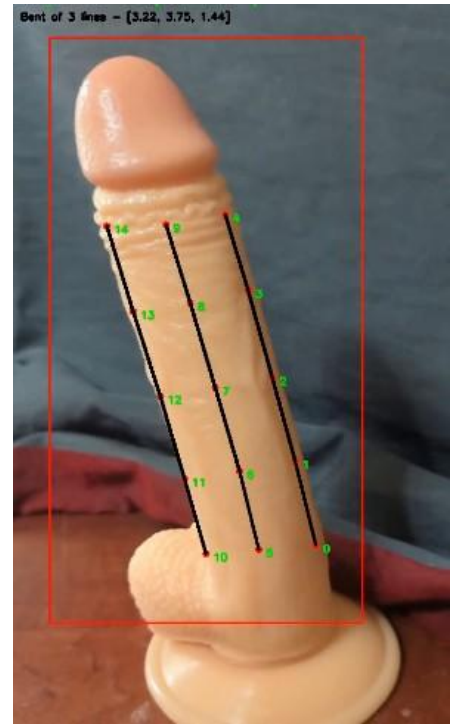

(a) Peyronie's penile image (Middle line angle: 67.51°)

(b) Normal Penile Image (Middle line angle: 3.75°)

Figure 6: Example of results

Architecture 2, which incorporates segmentation, demonstrated significantly improved performance in accurately distinguishing between Peyronie's disease and normal anatomical variations, achieving higher sensitivity, specificity, and overall diagnostic reliability.

## 5 Conclusion

This study presents an innovative AI-driven tool for diagnosing Peyronie's disease (PD) using advanced computer vision techniques to measure penile curvature angles through keypoint detection. Our methodology effectively addresses the limitations of traditional diagnostic methods, which often involve subjective assessments or invasive procedures. The approach demonstrates high precision and reliability, achieving a sensitivity of 96.7% and a specificity of 100% in distinguishing between Peyronie's disease and normal anatomical variations.

By enabling non-invasive, accurate assessments, our tool enhances the diagnostic process, providing healthcare providers with a more convenient and effective method for evaluating PD. The integration of video-based analysis allows for comprehensive evaluations, accommodating the natural variability in anatomical presentation and improving diagnostic accuracy.

Overall, this AI-driven approach represents a significant advancement in urology, offering a practical solution that enhances patient comfort and care. Future research may explore broader clinical applications and the integration of this technology with other diagnostic tools to further improve its utility and impact in healthcare settings.